\documentclass[aps,letter,preprint,amsfonts,amsmath,amssymb,prb]{revtex4}
\usepackage{amsmath,amsfonts,amssymb,bm}
\usepackage{array,multirow,dcolumn}
\usepackage{graphicx}
\usepackage{graphics}
\usepackage{color}
\begin{document}
\title{Diffusion coefficient and shear viscosity of rigid water models}

\author{Sami Tazi$^1$, Alexandru Bo\c{t}an$^{1,2}$, Mathieu Salanne$^1$, Virginie
Marry$^1$, Pierre Turq$^1$ and Benjamin Rotenberg$^1$}

\affiliation{$^1$ CNRS and UPMC Univ. Paris-06, 
UMR 7195 PECSA, 75005 Paris, France}
\affiliation{$^2$ IFP Energies nouvelles, 
1-4 av. de Bois Pr\'eau, 92852 Rueil-Malmaison, France}

\begin{abstract}
We report the diffusion coefficient and viscosity of popular
rigid water models: Two non-polarizable ones (SPC/E with 3 sites, 
and TIP4P/2005 with 4 sites) and a polarizable one (Dang-Chang,
4 sites). We exploit the dependence of the diffusion coefficient
on the system size 
[Yeh and Hummer, \textit{J. Phys. Chem. B} \textbf{108}, 15873 (2004)] 
to obtain the size-independent value. This also provides an
estimate of the viscosity of all water models, which we compare
to the Green-Kubo result. In all cases, a good agreement is found.
The TIP4P/2005 model is in better agreement with the
experimental data for both diffusion and viscosity.
The SPC/E and Dang-Chang water overestimate the diffusion
coefficient and underestimate the viscosity.
\end{abstract}
\maketitle

\section{Introduction}

The primary importance of water in biological, environmental or
technical processes explains the enormous efforts devoted 
to its modelling on the microscopic scale.
Many classical force fields for molecular simulations
have been introduced to capture different aspects of its complex
behaviour, including a number of "anomalies".
These models vary in the number of considered interaction sites,
their flexibility or rigidity and their treatment or neglect of polarizability.
Reviews of water models~\cite{guillot2002a,vega2008a,vega2011a}
have addressed their relative merits to model the thermodynamic
(e.g. phase diagram or heat capacities), 
structural and dynamic properties of water.
The latter usually include the diffusion coefficient and 
the shear viscosity of the model, which are key quantities
in determining the transport of water and solutions
at rest or under hydrodynamic flows.
In their recent review of the properties of rigid water models,
Vega and Abascal underlined that dynamic properties are very sensitive 
to the details of the water model and that "the diffusion coefficient
has not received the attention it should deserve as a target property"
in developing potential models~\cite{vega2011a}.

The computation of diffusion coefficients from molecular dynamics simulation
must be done carefully, as the result may depend on the simulation parameters
(cut-off radius or method to compute electrostatic interactions), 
but also exhibits a systematic size-dependence
arising from the screening of hydrodynamic flows under periodic boundary
conditions~\cite{dunweg1993a,yeh2004a}. When deriving the scaling of
the diffusion coefficient with the system size, which involves the
viscosity of the fluid, Yeh and Hummer investigated the 
TIP3P~\cite{jorgensen1983a} water model commonly used
for biomolecular simulations. In the present work, we investigate the diffusion
coefficient and viscosity of three popular rigid water models of different complexity. 
The SPC/E model~\cite{berendsen1987a}, involves three interaction sites
and is routinely used for the simulation of liquid water 
and electrolyte solutions.
The TIP4P/2005 model~\cite{abascal2005a}, with four interaction sites,
is usually the best rigid non-polarizable model for the description
of the phase diagram and the physical properties of bulk solid 
and liquid water.
The Dang-Chang model~\cite{dang1997a} is a four-site polarizable 
model, which allows some transferability from water clusters to 
liquid water and the description of the water-air interface.
For these three models, the system size-dependence of the diffusion
coefficient allows to determine the size-independent value
and provides an estimate of the viscosity. We also estimate the latter
from the Green-Kubo relation.

\section{Methods}

\begin{table}[htb]
\begin{center}
\begin{tabular}{lccccccc}
Model & $d_{\rm OH}$ (\AA)  & $d_{\rm OM}$ (\AA) & angle ($^\circ$) &
$\epsilon_{\rm O}$ (kcal/mol) & $\sigma_{\rm O}$ (\AA) &  
$q_{\rm H}$  & $\alpha_{\rm M}$ (\AA$^3$) \\\hline
SPC/E      & 1      & -      & 109.471 & 0.1554 & 3.1656 & 0.4238 &   -     \\
TIP4P/2005 & 0.9572 & 0.1546 & 104.52  & 0.1852 & 3.1589 & 0.5564 &  -      \\
Dang-Chang & 0.9572 & 0.215  & 104.52  & 0.1825 & 3.2340 & 0.5190 &  1.444 \\\hline
\end{tabular}
\caption{\label{tab:models} Geometry and parameters defining the three water
models: Bond length $d_{\rm OH}$, position of the extra M site along the
bissector $d_{\rm OM}$, angle between the OH bonds, Lennard-Jones
parameters on the O atom $\epsilon_{\rm O}$ and $\sigma_{\rm O}$,
charge on the H atoms ($q_{\rm O}=-2q_{\rm H}$ for SPC/E,
$q_{\rm M}=-2q_{\rm H}$ for the other models) and polarizability of
the M site.}
\end{center}
\end{table}

The parameters defining the geometry and force field for the
three water models are summarized in Table~\ref{tab:models}.
The diffusion coefficient are computed using from 
the mean-squared displacement, using the Einstein relation :
\begin{equation}
D_{PBC}=\lim\limits_{t \to \infty} \frac{1}{6} 
\frac{ {\rm d} \left< |{\bf r}(t)-{\bf r}(0)|^2\right>}{{\rm d}t}
\end{equation}
The "PBC" subscript emphasizes the fact that the use of periodic boundary
conditions induces a box length dependence of the measured diffusion
coefficient as~\cite{dunweg1993a,yeh2004a}:
\begin{equation}
\label{eq:DvsL}
D_{PBC} = D_0 - \frac{2.837 k_BT}{6\pi \eta L}
\end{equation}
where $\eta$ is the shear viscosity of the solvent.
In practice a linear fit of $D_{PBC}$ vs $1/L$
provides both $D_0$ and an estimate of the 
viscosity $\eta_{PBC}$.
A more usual determination of the viscosity consists in
using the Green-Kubo relation: 
\begin{equation}
\label{eq:greenkubo}
\eta_{GK} = \frac{V}{k_BT} 
\displaystyle\int_0^\infty \langle \sigma_{\alpha\beta}(t)
\sigma_{\alpha\beta}(0) \rangle \ {\rm d}t
\end{equation}
involving the auto-correlation function (ACF) of
the off-diagonal components of the stress tensor $\sigma_{\alpha\beta}$.

We performed molecular dynamics simulations in the $NVT$ ensemble,
for $N$ ranging from 64 to 4096 for SPC/E,
2048 for TIP4P/2005 and 512 for Dang-Chang (DC). The volume
is set to ensure a density of 0.998~g.cm$^{-3}$,
and a Nose-Hoover thermostat is used to maintain a temperature
$T=300$~K. Long-range interactions are computed using 
the Ewald summation technique~\cite{ewald1921a} and a cut-off
is used for short-range interactions.
For the DC model, the induced dipoles are computed at each time
step by minimizing the polarization energy and a convergence
criterium of 10$^{-6}$ is used.
Simulations of 1 to 10~ns are performed using a time step of 1~fs
and positions are sampled every 100~fs.
The equations of motion for the rigid water molecules are
integrated using the SHAKE algorithm~\cite{ryckaert1977a}.
For simulations with the non-polarizable SPC/E and TIP4P/2005 models
we use the LAMMPS~\cite{LAMMPS} code. We also compared our results
for TIP4P/2005, which contains a massless site, with that obtained
using Fincham's implicit quaternion algorithm~\cite{fincham1992a} instead of SHAKE,
and the DLPOLY code~\cite{DLPOLY}. Since the same results were found,
we only report the ones obtained with SHAKE.
Simulations with the polarizable Dang-Chang model were performed
with FIST, the classical part of the CP2K simulation package~\cite{CP2K}.  
The diffusion coefficients are computed from the slope
of the mean-square displacement in the 100-300~ps window.
Uncertainties on the diffusion coefficients for a given system size
are estimated using the block averaging method~\cite{Frenkel}.
The uncertainty on the size-independent diffusion coefficient $D_0$
and viscosity estimate $\eta_{PBC}$ are obtained from the 
least-square fit of the size-dependent ones to Eq.~\ref{eq:DvsL}. 
For the Green-Kubo calculations we use the
systems with $N=256$ molecules and collect the components
of the stress tensor at every time step.
The ACF is averaged over the off-diagonal components,
and the viscosity is computed as the integral over 5~ps 
(see below). The uncertainty is estimated by computing the standard
deviation of the values obtained for the different components.

\section{Results}

\begin{figure}[ht!]
\centering
\includegraphics[width=8cm]{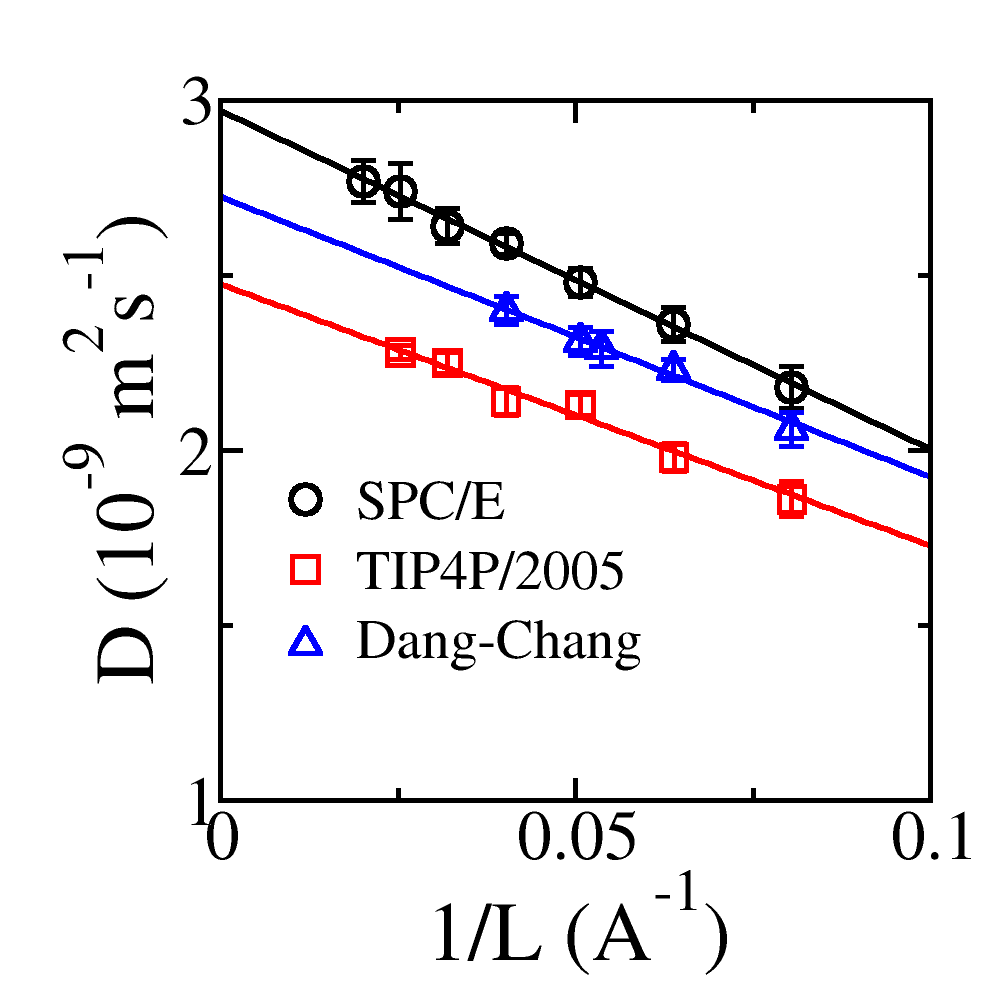}
\caption{Diffusion coefficient as a function
of the inverse box size. The size-independent diffusion coefficient
$D_0$ is the extrapolation at $1/L\to0$. The slope provides an
estimate of the viscosity $\eta_{PBC}$ using Eq.~\ref{eq:DvsL}.}
\label{fig:diff}
\end{figure}

The results for the diffusion coefficients as a function of system
size are reported in Figure~\ref{fig:diff}. For the three water models,
the scaling predicted by Eq.~\ref{eq:DvsL} is observed.
The corresponding size-independent diffusion coefficients $D_0$
and viscosities $\eta_{PBC}$ are summarized in Table~\ref{tab:results}.
All the considered models overestimate the diffusion coefficient
compared to the experimental result.
This is consistent for SPC/E with the result of Kerisit and 
Liu~\cite{kerisit2010a}. To the best of our knowledge, the
size-independent diffusion coefficient of the TIP4P/2005 and 
DC models had not been previously reported.
This fact was noticed as a caveat in Ref.~\cite{vega2011a},
where it was anticipated that extrapolating to infinite system
size would bring the result for this model, which is the only
rigid one to undersetimate the diffusion coefficient for typical
box sizes, closer to the experimental one. 
The extrapolated value $D_0$ turns out to be approximately
10\% larger. Nevertheless,
TIP4P/2005 provides the best agreement with the experimental value. 
This confirms the conclusion drawn, among rigid non-polarizable
models, from diffusion coefficients without taking into account
the system size-dependency~\cite{vega2011a}.
In their paper introducing Eq.~\ref{eq:DvsL}, 
Yeh and Hummer investigated the TIP3P water model~\cite{jorgensen1983a} 
and obtained $D_0=6.05~10^{-9}$~m$^2$s$^{-1}$~\cite{yeh2004a},
in even worse agreement with the experimental results.
The polarizable DC model overstimates the diffusion coefficient
by less than 20\% and performs better than SPC/E (30\%).

\begin{table}[!ht]
\begin{center}
\begin{tabular}{lccc}
\hline
Model & $D_0$ (10$^{-9}$~m$^2$s$^{-1}$) & $\eta_{PBC}$ (cP) &
$\eta_{GK}$ (cP)  \\ 
\hline
SPC/E       & 2.97$\pm$0.05    & 0.64$\pm$0.02    & 0.68$\pm$0.02  \\
TIP4P/2005  & 2.49$\pm$0.06    & 0.83$\pm$0.07    & 0.83$\pm$0.05  \\
Dang-Chang  & 2.72$\pm$0.09    & 0.78$\pm$0.06    & 0.74$\pm$0.09  \\
\hline
Exp.	    & 2.3$^a$   & \multicolumn{2}{c}{0.896$^b$}  \\
\hline
\end{tabular}
\end{center}
\caption{ \label{tab:results}
Size-independent diffusion coefficient $D_0$ and viscosity ($\eta_{PBC}$ 
from Eq.~\ref{eq:DvsL} and $\eta_{GK}$ from Eq.~\ref{eq:greenkubo}) 
of the three water models.
a: Reference~\cite{krynicki1978a}, b: Reference~\cite{harris2004a}.
}
\end{table}

\begin{figure}[ht!]
\centering
\includegraphics[height=7cm]{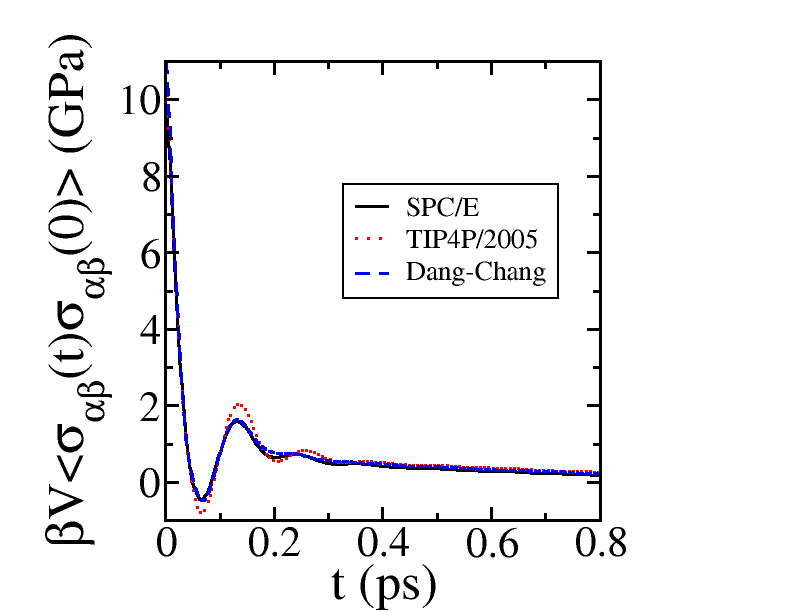}
\includegraphics[height=7cm]{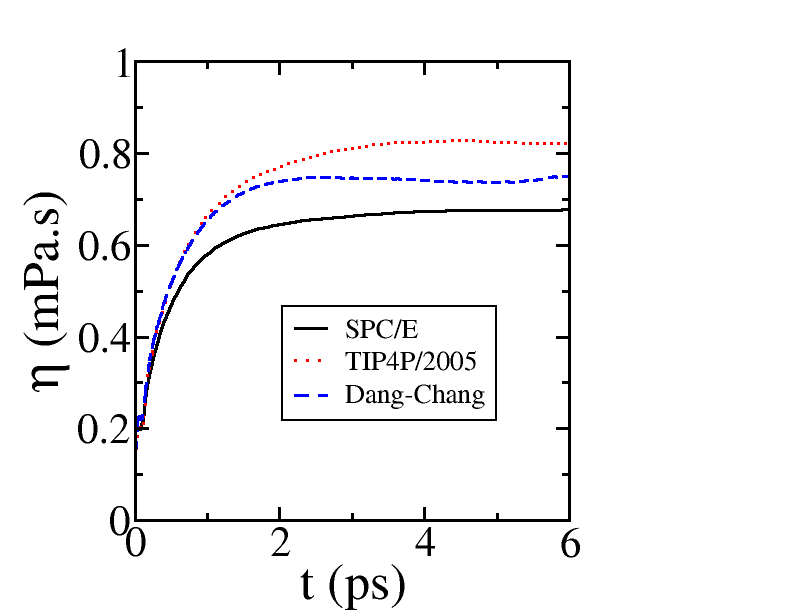}
\caption{Stress auto-correlation function and viscosity (see the Green-Kubo
equation~\ref{eq:greenkubo}, with $\beta=1/k_BT$) for the three water models.}
\label{fig:greenkubo}
\end{figure}

Before commenting the viscosity estimates $\eta_{PBC}$, 
we now turn to the Green-Kubo calculations.
Chen and Smit have recently shown that the viscosity
can be computed from equilibrium simulations with an accuracy
comparable to that obtained with non-equilibrium methods,
provided that the stress ACF is propertly sampled at short times
and that the integral Eq.~\ref{eq:greenkubo} is estimated
from times of the order of a few ps~\cite{chen2009a}.
Figure~\ref{fig:greenkubo} reports the stress ACF
at very short times and the time-dependent viscosity
(running integral of the ACF) for the three water models.
For SPC/E and TIP4P/2005, the ACF is very similar to the
one reported by Gonzalez and Abascal~\cite{gonzalez2010a},
with a fast decay and oscillations within a few 100~fs
which are more pronounced in the TIP4P/2005 case,
followed by a slower decay which contributes significantly to
the integral. The same behaviour is observed for the DC water
model, for which no such study had been yet reported.
As can be seen on Figure~\ref{fig:greenkubo},
the choice of an upper limit of 5~ps provides a good estimate
of the integral Eq.~\ref{eq:greenkubo}.

The viscosities $\eta_{GK}$ obtained by the Green-Kubo
formula and $\eta_{PBC}$ obtained from Eq.~\ref{eq:DvsL}
are summarized in Table~\ref{tab:results}.
For all three models, a good agreement between the two
approaches is obtained. 
The values for the SPC/E
model are comparable to most of the previously reported
values, both with equilibrium and non-equilibrium methods: 
0.64~cP~\cite{hess2002a}, 0.65~cP~\cite{guo2001a} and 0.67~cP~\cite{chen2009a}. 
A recent study of water confined in clay nanopores of
width larger than 4~nm concluded from non-equilibrium molecular
dynamics simulations to a viscosity of SPC/E water comparable
to the bulk value of 0.68-0.70~cP~\cite{botan2011a}.
Slightly larger values have also been reported for bulk SPC/E: 
0.72~cP~\cite{wu2006a} and 0.73~cP~\cite{gonzalez2010a}.
The value found for TIP4P/2005 is slightly smaller than
the only reported values in the literature, namely 
0.855~cP~\cite{gonzalez2010a} and 0.89~cP~\cite{guevara2011a}.
Nevertheless, we find an excellent agreement between
the two methods ($\eta_{PBC}$ and $\eta_{GK}$).
Since in both references the reported values for SPC/E
are also on the larger side (0.73 and 0.82~cP, respectively),
it is not entirely surprising. 
To the best of our knowledge, no viscosity of the 
DC model had been previously reported.
As for the size-independent diffusion coefficient,
TIP4P/2005 provides the best agreement with the experimental
viscosity, and the DC model performs better than SPC/E.
From the scaling of the diffusion coefficient with system
size, Yeh and Hummer had found for TIP3P a value of
$\eta_{PBC}=0.31\pm0.01$~cP~\cite{yeh2004a}, in worse
agreement with the experimental result, as for the diffusion
coefficient.

\section{Conclusion}

We have determined the size-independent diffusion coefficient 
and shear viscosity of three popular rigid water models: 
SPC/E, TIP4P/2005 and Dang-Chang.
A good agreement is found for each model between the
viscosity determined from the slope of $D$ vs the inverse box
size and from the Green-Kubo expression.
The three considered models overestimate the diffusion
coefficient and underestimate the viscosity.
The TIP4P/2005 model is in better agreement with the
experimental data for both diffusion and viscosity,
followed by the Dang-Chang and SPC/E models.

This confirms the success of TIP4P/2005 to 
model not only a large variety of thermodynamic and
material properties of bulk water, but also dynamic ones.
While it had been shown to correctly predict the
viscosity of liquid water at ambient conditions, as
well as as a function of temperature and 
pressure~\cite{vega2011a,guevara2011a},
the size-independent diffusion coefficient had not
been reported previously.
Although the prediction of the more complex (and thus
computationally more costly) Dang-Chang model are in slightly
worse agreement with the experimental data,
one should keep in mind that introducing polarizability 
should allow for a better transferability to other conditions, 
in particular liquid-solid interfaces and ionic solutions. 
The parametrization of a force field compatible with the Dang-Chang
water model to such situations is currently in progress~\cite{tazi2011a}.

\acknowledgements
The authors acknowledge financial support  from the Agence Nationale de la
Recherche under grant ANR-09-SYSC-012 and from IFP Energies Nouvelles. 

\vskip 1cm


\begin{thebibliography}{10}

\bibitem{guillot2002a}
Bertrand Guillot.
\newblock A reappraisal of what we have learnt during three decades of computer
  simulations on water.
\newblock {\em J. Mol. Liq.}, 101(1-3):219--260, 2002.

\bibitem{vega2008a}
Carlos Vega, Jose L.~F. Abascal, M.~M. Conde, and J.~L. Aragones.
\newblock What ice can teach us about water interactions: a critical comparison
  of the performance of different water models.
\newblock {\em Faraday Discuss.}, 141:251, 2008.

\bibitem{vega2011a}
Carlos Vega and Jose L.~F. Abascal.
\newblock Simulating water with rigid non-polarizable models: a general
  perspective.
\newblock {\em Phys. Chem. Chem. Phys.}, 2011.

\bibitem{dunweg1993a}
Burkhard D\"unweg and Kurt Kremer.
\newblock Molecular dynamics simulation of a polymer chain in solution.
\newblock {\em J. Chem. Phys.}, 99:6983, 2004.

\bibitem{yeh2004a}
I.-C. Yeh and G.~Hummer.
\newblock System-size dependence of diffusion coefficients and viscosities from
  molecular dynamics simulations with periodic boundary conditions.
\newblock {\em J. Phys. Chem. B}, 108(40):15873--15879, 2004.

\bibitem{jorgensen1983a}
William~L. Jorgensen, Jayaraman Chandrasekhar, Jeffry~D. Madura, Roger~W.
  Impey, and Michael~L. Klein.
\newblock Comparison of simple potential functions for simulating liquid water.
\newblock {\em J. Chem. Phys.}, 79:926, 1983.

\bibitem{berendsen1987a}
H.~J.~C. Berendsen, J.~R. Grigera, and T.~P. Straatsma.
\newblock The missing term in effective pair potentials.
\newblock {\em J. Phys. Chem.}, 91:6269--6271, 1987.

\bibitem{abascal2005a}
Jose L.~F. Abascal and Carlos Vega.
\newblock A general purpose model for the condensed phases of water:
  {TIP4P/2005}.
\newblock {\em J. Chem. Phys.}, 123:234505, 2005.

\bibitem{dang1997a}
Liem~X. Dang and Tsun-Mei Chang.
\newblock Molecular dynamics study of water clusters, liquid, and liquid--vapor
  interface of water with many-body potentials.
\newblock {\em J. Chem. Phys.}, 106:8149, 1997.

\bibitem{ewald1921a}
Paul~P. Ewald.
\newblock The calculation of optical and electrostatic lattice potentials.
\newblock {\em Ann. Phys.}, 64:253, 1921.

\bibitem{ryckaert1977a}
J.-P. Ryckaert, G.~Ciccotti, and H.~J.~C. Berendsen.
\newblock Numerical integration of the cartesian equations of motion of a
  system with constraints: molecular dynamics of $n-$alkanes.
\newblock {\em J. Comput. Phys.}, 23(3):327--341, 1977.

\bibitem{LAMMPS}
{LAMMPS}.
\newblock http://lammps.sandia.gov.

\bibitem{fincham1992a}
David Fincham.
\newblock Leapfrog rotational algorithms.
\newblock {\em Molec. Phys.}, 8(3-5):165--178, 1992.

\bibitem{DLPOLY}
{DLPOLY}.
\newblock http://www.ccp5.ac.uk/DL POLY.

\bibitem{CP2K}
{CP2K}.
\newblock http://cp2k.berlios.de, 2004.

\bibitem{Frenkel}
Daan Frenkel and Berend Smit.
\newblock {\em Understanding Molecular Simulations, From Algorithms to
  Applications}.
\newblock Academic Press, 2002.

\bibitem{kerisit2010a}
Sebastien Kerisit and Chongxuan Liu.
\newblock Molecular simulation of the diffusion of uranyl carbonate species in
  aqueous solution.
\newblock {\em Geochim. Cosmochim. Acta}, 74:4937--4952, 2010.

\bibitem{krynicki1978a}
K.~Krynicki, C.D. Green, and D.W. Sawyer.
\newblock Pressure and temperature dependence of self-diffusion in water.
\newblock {\em Faraday Discuss.}, 66:199--208, 1978.

\bibitem{harris2004a}
Kenneth~R. Harris and Lawrence~A. Woolf.
\newblock Temperature and volume dependence of the viscosity of water and heavy
  water at low temperatures.
\newblock {\em J. Chem. Eng. Data}, 49:1064--1069, 2004.

\bibitem{chen2009a}
Ting Chen, Berend Smit, and Alexis~T. Bell.
\newblock Are pressure fluctuation-based equilibrium methods really worse than
  nonequilibrium methods for calculating viscosities?
\newblock {\em J. Chem. Phys.}, 131:246101, 2009.

\bibitem{gonzalez2010a}
Miguel~Angel Gonzalez and Jose~L.F. Abascal.
\newblock The shear viscosity of rigid water models.
\newblock {\em J. Chem. Phys.}, 132:096101, 2010.

\bibitem{hess2002a}
Berk Hess.
\newblock Determining the shear viscosity of model liquids from molecular
  dynamics simulations.
\newblock {\em J. Chem. Phys.}, 116:209, 2002.

\bibitem{guo2001a}
Guang-Jun Guo and Yi-Gang Zhang.
\newblock Equilibrium molecular dynamics calculation of the bulk viscosity of
  liquid water.
\newblock {\em Molec. Phys.}, 99:283--289, 2001.

\bibitem{botan2011a}
Alexandru Bo\c{t}an, Benjamin Rotenberg, Virginie Marry, Pierre Turq, and
  Beno\^it Noetinger.
\newblock Hydrodynamics in clay nanopores.
\newblock {\em J. Phys. Chem. C}, 115(32):16109--16115, 2011.

\bibitem{wu2006a}
Yujie Wu, Harald~L. Tepper, and Gregory~A. Voth.
\newblock Flexible simple point-charge water model with improved liquid-state
  properties.
\newblock {\em J. Chem. Phys.}, 124:024503, 2006.

\bibitem{guevara2011a}
Gabriela Guevara-Carrion, Jadran Vrabec, and Hans Hasse.
\newblock Prediction of self-diffusion coefficient and shear viscosity of water
  and its binary mixtures with methanol and ethanol by molecular simulation.
\newblock {\em J. Chem. Phys.}, 134:074508, 2011.

\bibitem{tazi2011a}
Sami Tazi, John Molina, Benjamin Rotenberg, Pierre Turq, Rodolphe Vuilleumier and Mathieu Salanne.
\newblock A transferable ab-initio based force field for aqueous ions.
\newblock {\em J. Chem. Phys.}, 136:114507, 2012.

\end{thebibliography}

\end{document}